# Site Occupancy and Lattice Parameters in Sigma-Phase Co-Cr alloys

J. Cieślak[1], S. M. Dubiel[1*] and M. Reissner[2]

[1] Faculty of Physics and Applied Computer Science, AGH University of Science and Technology, PL-30-059 Kraków, Poland

[2] Institute of Solid State Physics, Vienna University of Technology, A-1040 Wien, Austria

**Abstract**

Neutron diffraction technique was used to study distribution of Co and Cr atoms over different lattice sites as well as lattice paramaters in sigma-phase $Co_{100-x}Cr_x$ compounds with $x$ = 57.0, 62.7 and 65.8. From the diffractograms recorded in the temperature range of 4.2 – 300 K it was found that all five sites A, B, C, D and E are populated by both kinds of atoms. Sites A and D are predominantly occupied by Co atoms while sites B, C and E by Cr atoms. The unit cell parameters $a$ and $c$, hence the unit cell volume, increase with $x$, the increase being characteristic of the lattice paramater and temperature. Both $a$ and $c$ show a non-linear increase with temperature.

Key words: Neutron diffraction, sigma phase, Co-Cr alloys

* Corresponding author: <dubiel@novell.ftj.agh.edu.pl>  (S. M. Dubiel)

## 1. Introduction

Among about 50 known examples of the sigma-phase in binary alloys, there is the Co-Cr one [1,2]. According to the binary-alloys phase diagram [3], the sigma-phase in this system can be formed within Cr concentration between 54 and 67 at% and a transformation of the original bcc-phase into the sigma phase can be done by isothermal annealing in the temperature interval of 600 – 1280$^o$C. The existence and properties of the sigma-phase in this alloy system are of a great scientific and technological interest. The former because the sigma phase in an $Co_{0.435}Cr_{0.565}$ alloy was reported, based on an indirect method, to be magnetic [4], and if so, this would be merely a third example of a magnetic sigma-phase known in the family of the binary alloy systems. The technological importance follows, among other, from the fact that Co-Cr alloys are used as dental and surgical alloys [5,6].

The paper is aimed at determination site populations by Co and Cr atoms in the sigma-phase of the Co-Cr system as well as lattice parameters. The sigma-phase, originally found and identified in the Fe-Cr system, has tetragonal structure. Its unit cell contains 30 atoms distributed over 5 non-equivalent crystallographic sites with a high (12-15) coordination number. For that reason, the sigma-phase is treated as a member of the so-called Frank-Kasper phases family. Its physical properties are charcateristic of a given alloy, but its common feature is hardness and brittleness. The latter makes a precipitation of the sigma-phase in various materials of technological importance e. g. stainless steels a very unwanted phenomenon as it causes significant deterioration of their useful mechanical properties.

The knowledge of the distribution of constituting atoms over the sites is important not only *per se* but also because it helps to properly interpret measurements performed with microscopic methods such as the nuclear magnetic resonance [7] or Mössbauer spectroscopy [8,9]. It is also very useful as input for performing theoretical calculations of the electronic structure of the sigma-phase [8,9].

In this study sigma-phase samples of $Co_{100-x}Cr_x$ with three different compositions were investigated by means of polycrystalline neutron diffraction (ND) techniques.

## 2. Experimental

Master alloys of $\alpha$-$Co_{100-x}Cr_x$ with $x$ = 57, 61 and 65 nominally, were prepared by melting appropriate amounts of Co (99.95% purity) and Cr (99.5 % purity) in an arc furnace under protective argon atmosphere. The melting process was repeated several times to ensure a better homogeneity of the alloys. The ingots received in that way, ca. 5 g, were next vacuum annealed at 1273 K for 6 days. Their chemical composition was determined on the annealed samples by electron probe microanalysis which gave x = 57.0, 62.7 and 65.8. For neutron diffraction measurements the samples were powdered by a mechanical attrition in an agate mortar.

The occupation numbers of particular sites and lattice parameters *a* and *c* of the unit cell were determined from neutron diffraction measurements performed at ILL Grenoble (D1A). The diffractograms obtained with λ = 1.91127 Å, which example is shown in Fig. 1, were measured in the temperature range between 4.2 K and 300 K. They were analyzed by means of Rietveld method (FULLPROF program) [11]. There were 22 refined parameters taken into account; 9 of them related to the background and position of the spectrum, next 9 parameters were connected with the scale parameter, line-widths and lattice constants, and 4 parameters were relevant to the Co/Cr occupation numbers of the five different lattice sites. In the calculations, the alloy concentration, *x*, was held fixed to the values obtained by the microprobe analysis, while the concentrations related to the particular sub-lattices were fitted. The inaccuracies of the occupation numbers were determined assuming they were mainly caused by the inhomogeneity of the samples as well as the limited accuracy of the determined composition, *x*. For that reason, for each spectrum the full fitting procedure was repeated for two additional compositions: *x+Δx* and *x-Δx* (*Δx* = 0.3 at% being the maximum expected error of the composition determination). The differences between the results of these two calculations, which did not exceed 0.5%, can be treated as the errors of the particular occupation numbers.

## 3. Results and their discussion

### 3. 1. Site occupation probability

In the following, only the average values are discussed. The data displayed in Fig. 2 give evidence that all five sites are occupied by both types of atoms constituting our samples. This observation is similar to that revealed earlier for the sigma-phase in Fe-Cr and Fe-V alloy systems [10]. The highest probability of finding Co atoms is for sites A and D. For the former it ranges between ~90% at x =57.0 and ~70% at x =65.8. The probability for site D is concentration independent and is equal to ~88%. Other sites i.e. B, C and E are mostly occupied by Cr atoms as the probabilities of finding Co atoms at these sites range between ~20% for x = 57.0 and ~10% for x = 65.8. It can be seen that, in general, there is no significant difference between the results determined from the diffractograms recorded at 4.2

K and 295 K. This can be considered as evidence that the analysis of the experimental data measured at two different temperatures is correct. On the other hand, the largest difference, namely of the order of 10% observed at $x = 57.0$ for site B and at $x = 65.8$ for site A, can be regarded as an upper limit of the accuracy of determining the probabilities in this experiment.

**3. 2. Lattice parameters**

Lattice parameters *a* and *c* of the unit cell as obtained from the diffractograms recorded at 4.2 K and 300 K are presented in Fig. 3 versus Cr concentration, *x*. As can be seen, at room temperature both of them increase with *x*, however, the increase is not linear and different for *a* and *c*. Namely, the value of the former increases from 8.775818 Å at $x = 57.0$ to 8.78865 Å for $x = 65.8$ i.e. the relative increase $\Delta a/a = 0.35\%$, whereas the value of the latter increases from 4.53363 Å at $x = 57.0$ to 4.53866 Å at $x = 65.8$ i.e. the relative increase $\Delta c/c = 0.11\%$ i.e. three-fold less. The *c/a* ratio changes from 0.5176 at $x = 57.0$ to 0.5164 at $x = 65.8$. The former figure is close to that given elsewhere [1], though the values of both lattice constants for that composition are in our case significantly smaller than those reported in [1].

The behavior of the lattice constants at 4.2 K is different: while *a* is only shifted downwards, on average by 0.0034 Å, *c* shows a shallow minimum at x = 62.7. The *c/a* ratio as determined for this temperature changes with growing *x* as follows: 0.5173, 0.5165, 0.5161. That means that the *c/a* ratio hardly depends on temperature.

Using the 300 K values of *a* and *c*, those of the unit cell volume, *V*, were calculated and are displayed in Fig. 4. For comparison, the *V*-values obtained for the sigma-phase $Fe_xCr_{100-x}$ and $Fe_xV_{100-x}$ compounds are also presented [12]. It is evident from that figure that all structure parameters are characteristic of a given alloy system and they show a quasi-linear composition dependence. It is also clear that the structural parameters are correlated with the atomic size of atoms constituting a given compound. In particular, the atom radius of Co is equal to 1.25 Å while that of Cr 1.30 Å.

Finally, the measurements of the diffractograms over the temperature interval of 4.2 –300 K enabled determination of the effect of temperature, *T*, on the lattice constants. As illustrated in Fig. 5 both *a* and *c* show a non-linear increase with *T*. It is also worth noticing that the lattice constant *a* shows a normal behaviour i.e. it is larger for a larger content of chromium at all temperatures measured, while *c* exhibits an anomaly viz. such correlation occurs only above ~250 K, while below this temperature its value for $x = 62.7$ becomes smaller than that for $x = 57.0$.

## 4. Conclusions

The results obtained and presented in this paper can be concluded as follows:

a) Co/Cr atoms are present on all five crystallographic sites; A and D sites are mostly populated by Co atoms with a probability between ~90 and ~70 %, whereas Cr atoms predominantly reside on sites B, C and E with the probability between ~20% and ~10% depending on the composition.

b) Lattice parameters $a$ and $c$ increase with the chromium content, $x$, the increase with temperature is different for both.

c) Lattice parameters $a$ and $c$ increase non-linearly with temperature.

d) Lattice constant $c$ shows some anomaly at $T \approx 250$ K for $x = 57.0$ and $x = 62.7$.

**Acknowledgements**

This study was carried out within a bilateral governmental Austro-Polish scientific co-operation – project WTZ 5/2009

**Figures**

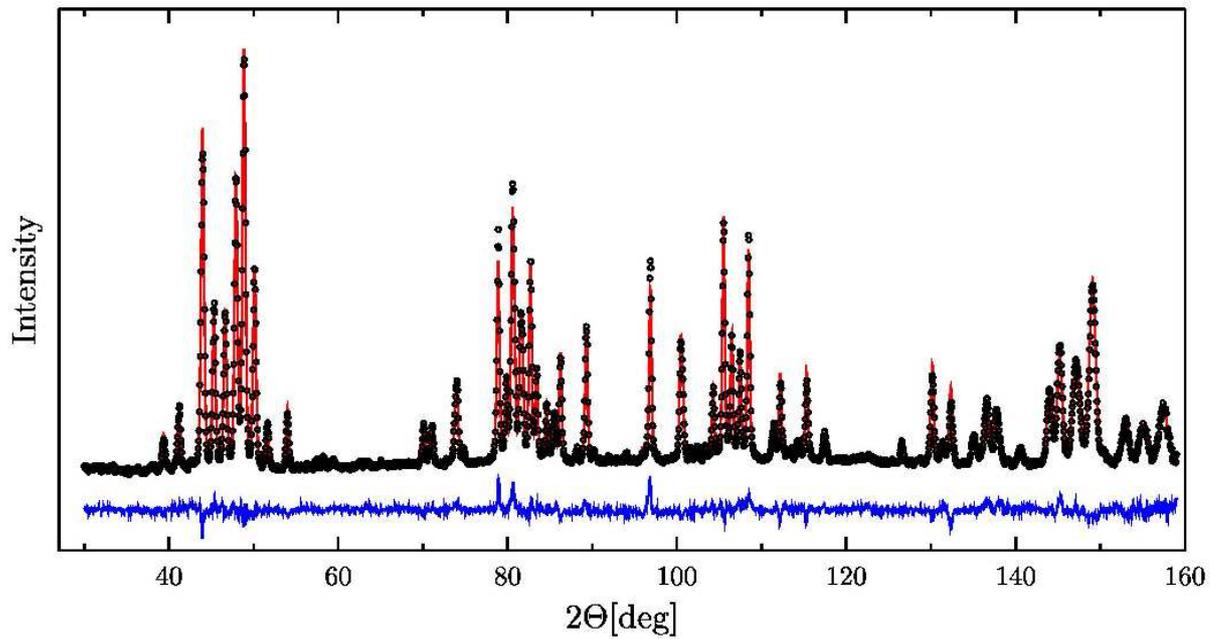

Fig. 1 Neutron diffractogram recorded at 300 K on the sigma-phase sample of $Co_{34.2}Cr_{65.8}$. The solid line stays for the best-fit obtained with the procedure described in the text. A difference diffractogram is marked, too.

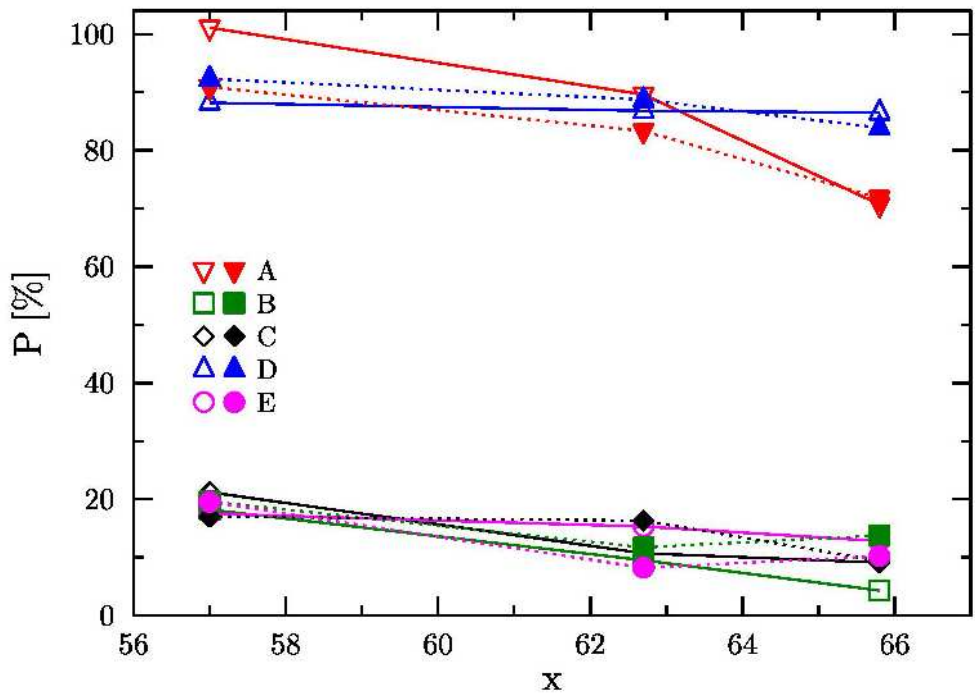

Fig. 2

Probability of finding Co atoms at different lattice sites in the sigma-phase $Co_{100-x}Cr_x$ compounds, *P*, versus chromium concentration, *x*. Solid lines connect the points obtained from the 300 K measurements while the dotted ones stay for the 4.2 K results. They are marked to guide the eyes.

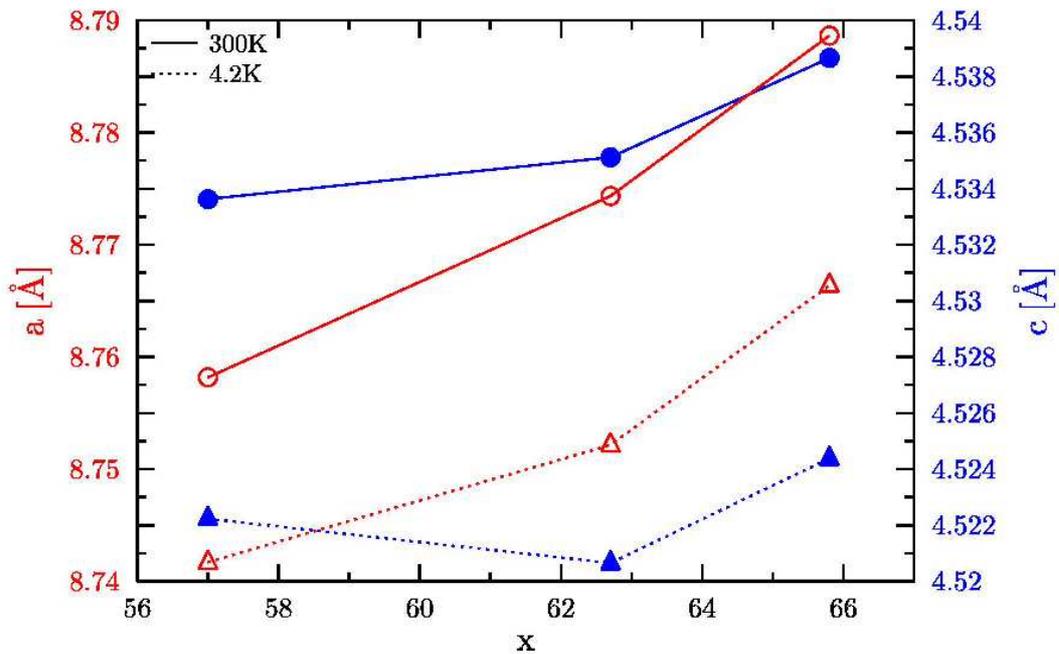

Fig. 3

Dependence of the lattice parameters *a* (open symbols) and *c* (full symbols) on chromium content, *x* as determined from the neutron diffractograms recorded at 4.2 and 300 K.

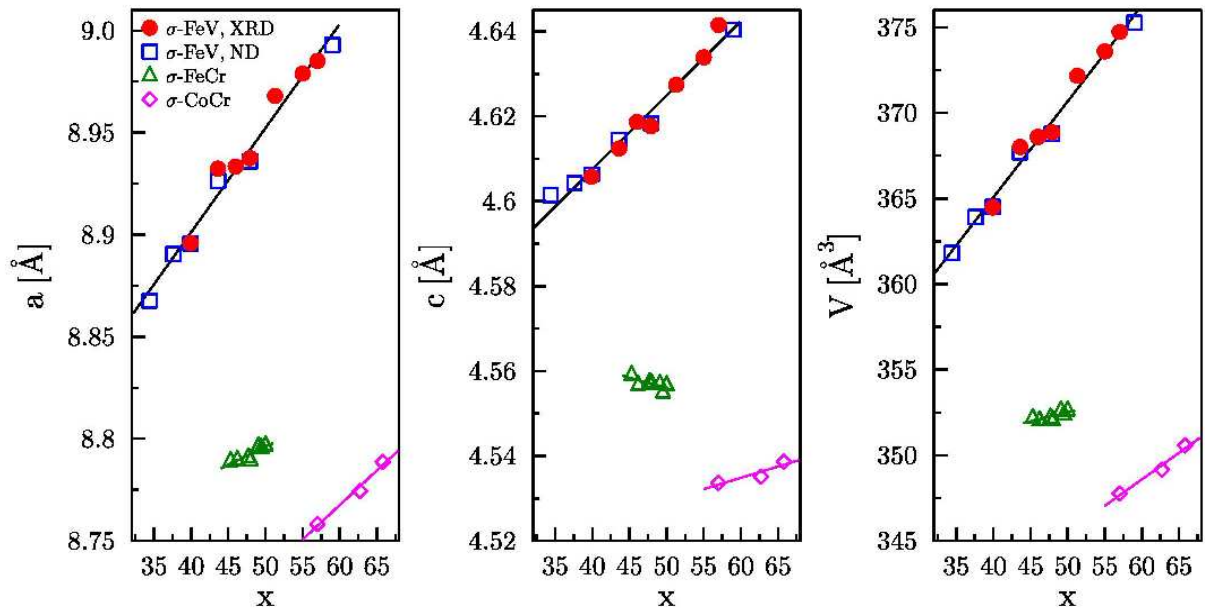

Fig. 4

Composition dependence of the lattice parameters *a* and *c*, and that of the unit cell volume, *V*, at 300 K for the sigma-phase in $Fe_{100-x}Cr_x$, $Fe_{100-x}V_x$ [12] and $Co_{100-x}Cr_x$ alloy systems.

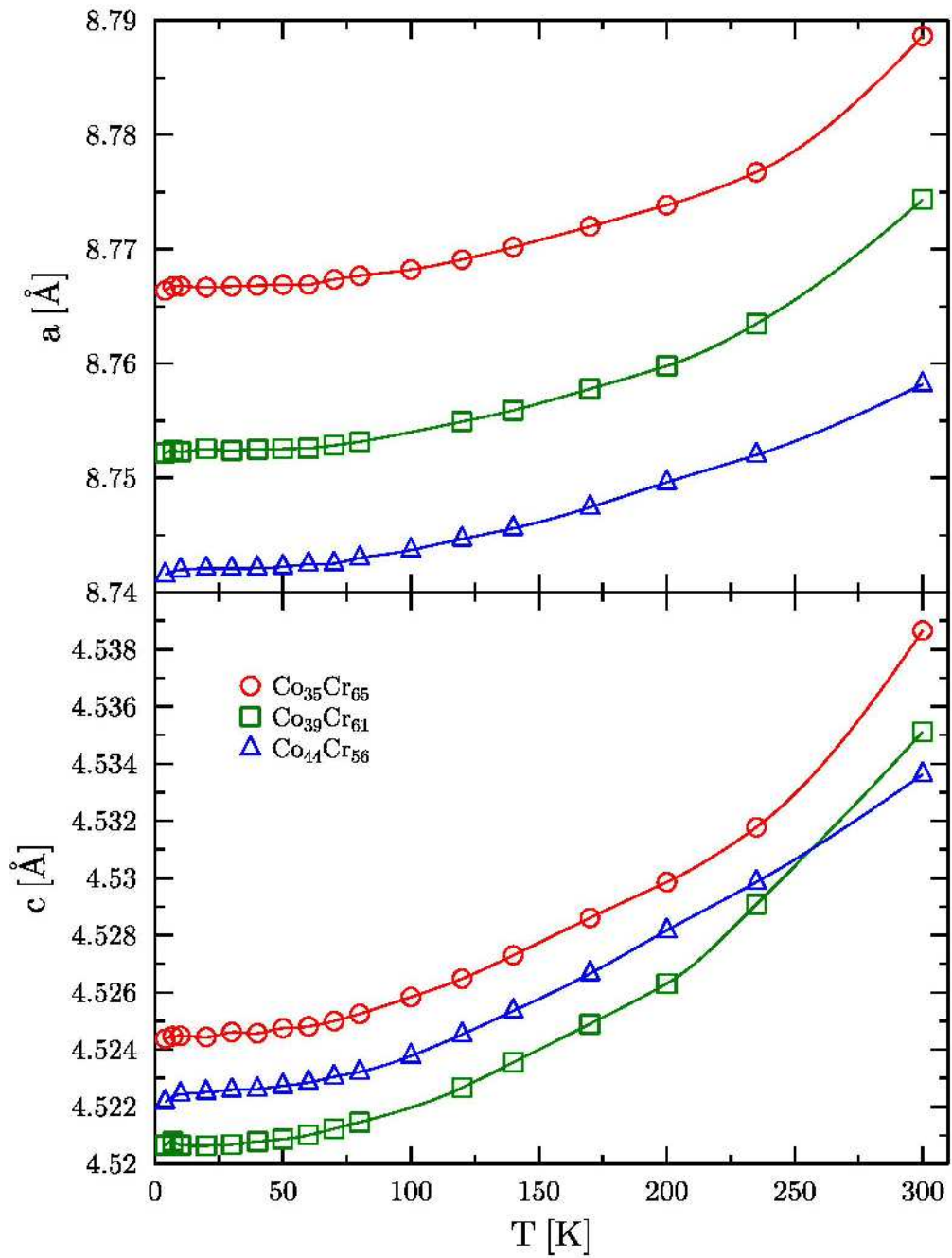

Fig. 5

Temperature dependence of the lattice parameters *a* and *c* for the sigma-phase Co-Cr samples. The solid lines are drawn to guide the eyes.